\documentclass[letterpaper, 11 pt, conference,one column]{ieeeconf}
\IEEEoverridecommandlockouts     
\usepackage{graphics} 
\usepackage{svg}
\usepackage{amsmath} 
\usepackage{mathtools}
\usepackage{amssymb}
\usepackage{float}
\usepackage{caption}
\usepackage{graphicx}
\usepackage[normalem]{ulem}
\usepackage{algorithm}
\usepackage{algorithmicx}
\usepackage{cite}
\usepackage{algpseudocode}
\usepackage[hang,flushmargin]{footmisc}

\usepackage{soul}
\usepackage{comment}
\usepackage{tikz} % For drawing custom dots
\usepackage{xcolor}
\usepackage{pgfplots}
\usepackage{pgfplotstable}
\usepackage{url}
\usepackage{graphviz}
\usetikzlibrary{math}
\usetikzlibrary{shapes,arrows}
\usetikzlibrary{arrows.meta}
\tikzset{>={Latex[width=2.5mm,length=2.5mm]}}
\usetikzlibrary{positioning}
\tikzstyle{block}=[draw opacity=0.7,line width=1.4cm]

\newcommand{\real}{{\mathbb{R}}}

 \newcommand{\boxend}{\hfill \ensuremath{\Box}}
\newtheorem{thm}{Theorem}[section]

\newtheorem{lem}{Lemma}[section]
\newtheorem{defn}{Definition}

\newcommand{\longthmtitle}[1]{\mbox{}\textit{{(#1):}}}

\allowdisplaybreaks
\title{\LARGE \bf
\textsf{ResQue Greedy}: Rewiring Sequential Greedy for \\Improved Submodular Maximization\\

}
\author{Joan Vendrell$^\dagger$, Alan Kuhnle$^\ddagger$ and Solmaz Kia$^\dagger$, \emph{Senior Member, IEEE} 
\thanks{$^\dagger$ The first and the third authors are with the Mechanical and Aerospace Engineering Department of University of California Irvine, Irvine, CA, USA
{\tt\small jvendrel,solmaz@uci.edu}.  $^\ddagger$ The second author is with the Computer Science and Engineering Department of the Texas A\& M University, Texas, TX, USA {\tt\small kuhnle@tamu.edu}.
    }
}
\allowdisplaybreaks
\newcolumntype{d}[1]{>{\centering\arraybackslash}m{#1\linewidth}}
\pdfoutput=1
\pgfplotsset{compat=1.18}
\begin{document}
\maketitle
\begin{abstract}
This paper introduces Rewired Sequential Greedy (\textsf{ResQue Greedy}), an enhanced approach for submodular maximization under cardinality constraints. By integrating a novel set curvature metric within a lattice-based framework, \textsf{ResQue Greedy} identifies and corrects suboptimal decisions made by the standard sequential greedy algorithm. Specifically, a curvature-aware rewiring strategy is employed to dynamically redirect the solution path, leading to improved approximation performance over the conventional sequential greedy algorithm without significantly increasing computational complexity. Numerical experiments demonstrate that \textsf{ResQue Greedy} achieves tighter near-optimality bounds compared to the traditional sequential greedy method.
\end{abstract}


%%%%%%%%%%%%%%%%%%%%%%%%%%%%%%%%%%%%%%%%%%%%%%%%%%%%%%%%%%%%%%%%%%%%%%%%%%%%%%%%
\section{Introduction}
\label{sec::intro}
This paper considers the problem of submodular maximization subject to a cardinality constraint. In this problem, the goal is to select up to $\kappa\in\mathbb{Z}_{>0}$ strategies from a discrete strategy set $\mathcal{P}$ in a way that a normal monotone increasing and submodular utility function \(f:2^\mathcal{P}\to\mathbb{R}^{+}\) is maximized. The optimization problem is expressed as
\begin{subequations}\label{eq::mainProblem}
\begin{align}
    &\underset{\mathcal{S}\in \mathcal{I}}{\textup{max}}\,f(\mathcal{S}) \\
    &\mathcal{I} = \big\{ \mathcal{S} \subset \mathcal{P}\,\big|\,\, |\mathcal{S}|\leq \kappa\big\},
    \label{eq::mainProblem_b}
\end{align}
\end{subequations}
\noindent where $\mathcal{I}$ is a \emph{uniform matroid}, restricting the number of strategies that can be chosen up to $\kappa$. The access to the utility function is via an oracle that returns \(f(\mathcal{S})\) for any set \(\mathcal{S} \subset \mathcal{P}\) (value oracle model). 

\noindent \begin{minipage}[b]{0.55\textwidth} Many resource allocation problems in applications such as sensor allocation~\cite{RL-NM-RH:23}, robot deployment~\cite{NR-SSK:21}, re-balancing in shared mobility systems~\cite{PGS-IB-AK-MK:21} or machine scheduling in manufacturing~\cite{SL:20} are cast as the submodular maximization  problem~\eqref{eq::mainProblem}. However, submodular function maximization problems such as~\eqref{eq::mainProblem} are often NP-hard~\cite{GLN-LAW-MLF:78}, meaning that optimal solution cannot be achieved in polynomial time. Fortunately, submodularity is a property of set functions with deep theoretical implications, which enables the establishment of constant factor approximate (suboptimal) solutions. It is well-established that for the problem~\eqref{eq::mainProblem}, the Sequential Greedy Algorithm, Algorithm~\ref{alg:sequential},
\end{minipage} \quad
\begin{minipage}[b]{0.43\textwidth}
\begin{algorithm}[H]
\caption{Sequential Greedy Algorithm}\label{alg:sequential}
\begin{algorithmic}[1] % Added the [1] for line numbers
\Require Ground set $\mathcal{P}$ and utility function $f$
\Ensure $\mathcal{S}^\text{SG}\subset \mathcal{P}$ satisfying $|\mathcal{S}^\text{SG}|\leq \kappa$
\State $\mathcal{S}_0 \gets \emptyset$
\For{$i \in \{0, \dots, \kappa-1\}$} % Standard \For loop
  \State $s_{i+1} \gets \text{argmax}_{s \in \mathcal{P} \backslash \mathcal{S}_{i}}{ \Delta_f(s|\mathcal{S}_{i})}\footnotemark$
  \State $\mathcal{S}_{i+1} \gets \mathcal{S}_{i} \cup \{s_{i+1}\}$
\EndFor
\State $\mathcal{S}^\text{SG} \gets \mathcal{S}_{\kappa}$
\end{algorithmic}
\end{algorithm}
\end{minipage}
\footnotetext{$\Delta_f(s|\mathcal{S}_{i-1})\!=\!f(\mathcal{S}_{i-1}\cup\{s\})\!-\!f(\mathcal{S}_{i-1})$, is referred to as \emph{marginal gain}.}
\noindent  \!\!can achieve an approximation ratio strictly better than $(1-\text{e}^{-1})$. The optimality gap of $(1-\text{e}^{-1})$ is the worst case bound that represents a fundamental theoretical limit on the achievable performance of polynomial algorithms. In practice however, the actual achieved bound can be better than $(1-\text{e}^{-1})$ on any specific instance of the problem. 

The hard bound of $(1-\text{e}^{-1})$ was established for submodular functions in general. However, by considering the curvature of the submodular function, which quantifies the rate of diminishing returns, improved approximation bounds are possible in terms of the curvature parameter.
\begin{defn}[Total curvature of a submodular function~\cite{MC-GC:84}]
    Let $f: 2^{\mathcal{P}} \rightarrow \mathbb{R}^{+}$ be a monotone submodular function over a ground set $\mathcal{P}$. The \emph{total curvature} $c$ of $f$ is defined as:
    \begin{equation}\label{eq::total_curv}c = 1 - \min_{\substack{e \in \mathcal{P} \setminus \mathcal{S} \\ \mathcal{S} \subset \mathcal{P}}} \frac{f(\mathcal{S} \cup \{e\}) - f(\mathcal{S})}{f(\{e\}) - f(\emptyset)}\end{equation}
    where $0 \leq c \leq 1$.\boxend
    \label{defn:curvature}
\end{defn}
The total curvature $c$ quantifies how much the returns may diminish, or how submodular the function is. When $c=0$, the function is modular (additive), a special case where an optimal solution can be found in polynomial time. When $c=1$, the function exhibits maximum submodularity, and no ratio better than $1-\text{e}^{-1}$ is possible. For $0 < c \le 1$, Conforti et al.~\cite{MC-GC:84} quantified the optimality gap for a solution generated by the sequential greedy algorithm as $\frac{1}{c}(1-\text{e}^{-c})$. 
\begin{figure*}[t]
    \centering
    \begin{minipage}{0.3\textwidth}
        \centering
        \begin{tikzpicture}[
            scale=0.35,
            every node/.style={draw, circle, inner sep=1pt, font=\tiny},
            level 1/.style={sibling distance=4cm},
            level 2/.style={sibling distance=1.5cm},
            level 3/.style={sibling distance=0.8cm}
        ]
        % Level 5
        \node (12345) at (0,12) {12345};
        % Level 4
        \node (1234) at (-4,9.5) {1234};
        \node (1235) at (-2,9.5) {1235};
        \node (1345) at (0,9.5) {1345};
        \node (1245) at (2,9.5) {1245};
        \node (2345) at (4,9.5) {2345};
        % Level 3
        \node (123) at (-6.8,6.5) {123};
        \node (124) at (-5.3,6.5) {124};
        \node (125) at (-3.8,6.5) {125};
        \node (134) at (-2.3,6.5) {134};
        \node (135) at (-0.8,6.5) {135};
        \node (145) at (0.7,6.5) {145};
        \node (234) at (2.2,6.5) {234};
        \node (235) at (3.7,6.5) {235};
        \node (245) at (5.2,6.5) {245};
        \node (345) at (6.7,6.5) {345};
        % Level 2
        \node (12) at (-6.7,4) {12};
        \node (13) at (-5.2,4) {13};
        \node (14) at (-3.7,4) {14};
        \node (15) at (-2.2,4) {15};
        \node (23) at (-0.7,4) {23};
        \node (24) at (0.7,4) {24};
        \node (25) at (2.2,4) {25};
        \node (34) at (3.7,4) {34};
        \node (35) at (5.2,4) {35};
        \node (45) at (6.7,4) {45};
        % Level 1
        \node (1) at (-4,2) {1};
        \node (2) at (-2,2) {2};
        \node (3) at (0,2) {3};
        \node (4) at (2,2) {4};
        \node (5) at (4,2) {5};
        % Level 0
        \node (0) at (0,0) {$\emptyset$};
        
        % Connections
        \foreach \i in {1,2,3,4,5}
            \draw[blue!90!white,dashed,thick] (0) -- (\i);
        \draw[black!20!white,dashed] (1) -- (12)
            (1) -- (13)
            (1) -- (14)
            (1) -- (15)
            (2) -- (12);
        \draw[blue!90!white,dashed,thick] (2) -- (23)
            (2) -- (24)
            (2) -- (25);
        \draw[black!20!white,dashed] (3) -- (13)
            (3) -- (23)
            (3) -- (34)
            (3) -- (35)
            (4) -- (14)
            (4) -- (24)
            (4) -- (34)
            (4) -- (45)
            (5) -- (15)
            (5) -- (25)
            (5) -- (35)
            (5) -- (45)
            (12) -- (123)
            (12) -- (124)
            (12) -- (125);
        \draw[black!20!white,dashed] (13) -- (123)
            (13) -- (134)
            (13) -- (135)
            (14) -- (124)
            (14) -- (134)
            (14) -- (145)
            (23) -- (234)
            (23) -- (123)
            (23) -- (235)
            (24) -- (234)
            (24) -- (124)
            (24) -- (245)
            (34) -- (134)
            (34) -- (234)
            (34) -- (345)
            (15) -- (125)
            (15) -- (145)
            (15) -- (135)
            (25) -- (245)
            (25) -- (235)
            (25) -- (125)
            (35) -- (135)
            (35) -- (235)
            (35) -- (345)
            (45) -- (145)
            (45) -- (245)
            (45) -- (345);
        \draw[black!20!white,dashed] (123) -- (1234)
            (123) -- (1235)
            (125) -- (1235)
            (125) -- (1245);
        \draw[black!20!white,dashed] (134) -- (1234)
            (134) -- (1345)
            (234) -- (1234)
            (234) -- (2345)
            (124) -- (1234)
            (124) -- (1245)
            (135) -- (1345)
            (135) -- (1235)
            (145) -- (1245)
            (145) -- (1345)
            (235) -- (1235)
            (235) -- (2345)
            (245) -- (1245)
            (245) -- (2345)
            (345) -- (2345)
            (345) -- (1345);
        \draw[black!20!white,dashed] (1234) -- (12345)
            (1235) -- (12345)
            (1345) -- (12345)
            (1245) -- (12345)
            (2345) -- (12345);
        \draw[blue!90!white,dashed,thick] (25) -- (245)
        (2) -- (12)
        (25) -- (125)
        (235) -- (1235);
        \draw[->,blue!50!black,very thick] (0) -- (2);
        \draw[->,blue!50!black,very thick] (2) -- (25);
        \draw[->,blue!50!black,very thick] (25) -- (235);
        \draw[->,blue!50!black,very thick] (235) -- (2345);
        \end{tikzpicture}
        \label{fig:lattice}
        
        \text{(a) Sequential Greedy Algorithm.}
    \end{minipage}
    \hfill
    \begin{minipage}{0.3\textwidth}
    \begin{tikzpicture}[
            scale=0.35,
            every node/.style={draw, circle, inner sep=1pt, font=\tiny},
            level 1/.style={sibling distance=4cm},
            level 2/.style={sibling distance=1.5cm},
            level 3/.style={sibling distance=0.8cm}
        ]
        % Level 5
        \node (12345) at (0,12) {12345};
        % Level 4
        \node (1234) at (-4,9.5) {1234};
        \node (1235) at (-2,9.5) {1235};
        \node (1345) at (0,9.5) {1345};
        \node (1245) at (2,9.5) {1245};
        \node (2345) at (4,9.5) {2345};
        % Level 3
        \node (123) at (-6.8,6.5) {123};
        \node (124) at (-5.3,6.5) {124};
        \node (125) at (-3.8,6.5) {125};
        \node (134) at (-2.3,6.5) {134};
        \node (135) at (-0.8,6.5) {135};
        \node (145) at (0.7,6.5) {145};
        \node (234) at (2.2,6.5) {234};
        \node (235) at (3.7,6.5) {235};
        \node (245) at (5.2,6.5) {245};
        \node (345) at (6.7,6.5) {345};
        % Level 2
        \node (12) at (-6.7,4) {12};
        \node (13) at (-5.2,4) {13};
        \node (14) at (-3.7,4) {14};
        \node (15) at (-2.2,4) {15};
        \node (23) at (-0.7,4) {23};
        \node (24) at (0.7,4) {24};
        \node (25) at (2.2,4) {25};
        \node (34) at (3.7,4) {34};
        \node (35) at (5.2,4) {35};
        \node (45) at (6.7,4) {45};
        % Level 1
        \node (1) at (-4,2) {1};
        \node (2) at (-2,2) {2};
        \node (3) at (0,2) {3};
        \node (4) at (2,2) {4};
        \node (5) at (4,2) {5};
        % Level 0
        \node (0) at (0,0) {$\emptyset$};
        
        % Connections
        \foreach \i in {1,2,3,4,5}
            \draw[black!40!white,dashed] (0) -- (\i);
        \draw[black!40!white,dashed] (1) -- (12)
            (1) -- (13)
            (1) -- (14)
            (1) -- (15)
            (2) -- (12)
            (2) -- (23)
            (2) -- (24)
            (2) -- (25)
            (3) -- (13)
            (3) -- (23)
            (3) -- (34)
            (3) -- (35)
            (4) -- (14)
            (4) -- (24)
            (4) -- (34)
            (4) -- (45)
            (5) -- (15)
            (5) -- (25)
            (5) -- (35)
            (5) -- (45);
        \draw[black!40!white,dashed] (12) -- (123)
            (12) -- (124)
            (12) -- (125)
            (13) -- (123)
            (13) -- (134)
            (13) -- (135)
            (14) -- (124)
            (14) -- (134)
            (14) -- (145)
            (23) -- (234)
            (23) -- (123)
            (23) -- (235)
            (24) -- (234)
            (24) -- (124)
            (24) -- (245)
            (34) -- (134)
            (34) -- (234)
            (34) -- (345)
            (15) -- (125)
            (15) -- (145)
            (15) -- (135)
            (25) -- (245)
            (25) -- (235)
            (25) -- (125)
            (35) -- (135)
            (35) -- (235)
            (35) -- (345)
            (45) -- (145)
            (45) -- (245)
            (45) -- (345);
        \draw[black!40!white,dashed] (123) -- (1234)
            (123) -- (1235)
            (124) -- (1234)
            (124) -- (1245)
            (134) -- (1234)
            (134) -- (1345)
            (234) -- (1234)
            (234) -- (2345)
            (125) -- (1235)
            (125) -- (1245)
            (135) -- (1345)
            (135) -- (1235)
            (145) -- (1245)
            (145) -- (1345)
            (235) -- (1235)
            (235) -- (2345)
            (245) -- (1245)
            (245) -- (2345)
            (345) -- (2345)
            (345) -- (1345);
        \draw[black!40!white,dashed] (1234) -- (12345)
            (1235) -- (12345)
            (1345) -- (12345)
            (1245) -- (12345)
            (2345) -- (12345);     
        \draw[green!50!black,very thick] (0) -- (1);
        \draw[green!50!black,very thick] (0) -- (3);
        \draw[green!50!black,very thick] (0) -- (4);
        \draw[green!50!black,very thick] (0) -- (2);
        \draw[green!50!black,very thick] (1) -- (13);
        \draw[green!50!black,very thick] (1) -- (14);
        \draw[green!50!black,very thick] (1) -- (12);
        \draw[green!50!black,very thick] (3) -- (13);
        \draw[green!50!black,very thick] (3) -- (34);
        \draw[green!50!black,very thick] (4) -- (34);
        \draw[green!50!black,very thick] (4) -- (14);
        \draw[green!50!black,very thick] (4) -- (24);
        \draw[green!50!black,very thick] (3) -- (23);
        \draw[green!50!black,very thick] (2) -- (12);
        \draw[green!50!black,very thick] (2) -- (24);
        \draw[green!50!black,very thick] (2) -- (23);
        \draw[green!50!black,very thick] (13) -- (123);
        \draw[green!50!black,very thick] (13) -- (134);
        \draw[green!50!black,very thick] (14) -- (124);
        \draw[green!50!black,very thick] (14) -- (134);
        \draw[green!50!black,very thick] (12) -- (123);
        \draw[green!50!black,very thick] (12) -- (124);
        \draw[green!50!black,very thick] (23) -- (123);
        \draw[green!50!black,very thick] (23) -- (234);
        \draw[green!50!black,very thick] (34) -- (234);
        \draw[green!50!black,very thick] (34) -- (134);
        \draw[green!50!black,very thick] (24) -- (234);
        \draw[green!50!black,very thick] (24) -- (124);
        \draw[green!50!black,very thick] (134) -- (1234);
        \draw[green!50!black,very thick] (123) -- (1234);
        \draw[green!50!black,very thick] (124) -- (1234);
        \draw[green!50!black,very thick] (124) -- (1234);
        \draw[green!50!black,very thick] (234) -- (1234);
        \end{tikzpicture}
        \text{(b) Paths to the optimal solution.}
        \label{fig::lattice_combined}
    \end{minipage}
    \hfill
    \begin{minipage}{0.3\textwidth}
        \centering
        \begin{tikzpicture}[
            scale=0.35,
            every node/.style={draw, circle, inner sep=1pt, font=\tiny},
            level 1/.style={sibling distance=4cm},
            level 2/.style={sibling distance=1.5cm},
            level 3/.style={sibling distance=0.8cm}
        ]
        % Level 5
        \node (12345) at (0,12) {12345};
        % Level 4
        \node (1234) at (-4,9.5) {1234};
        \node (1235) at (-2,9.5) {1235};
        \node (1345) at (0,9.5) {1345};
        \node (1245) at (2,9.5) {1245};
        \node (2345) at (4,9.5) {2345};
        % Level 3
        \node (123) at (-6.8,6.5) {123};
        \node (124) at (-5.3,6.5) {124};
        \node (125) at (-3.8,6.5) {125};
        \node (134) at (-2.3,6.5) {134};
        \node (135) at (-0.8,6.5) {135};
        \node (145) at (0.7,6.5) {145};
        \node (234) at (2.2,6.5) {234};
        \node (235) at (3.7,6.5) {235};
        \node (245) at (5.2,6.5) {245};
        \node (345) at (6.7,6.5) {345};
        % Level 2
        \node (12) at (-6.7,4) {12};
        \node (13) at (-5.2,4) {13};
        \node (14) at (-3.7,4) {14};
        \node (15) at (-2.2,4) {15};
        \node (23) at (-0.7,4) {23};
        \node (24) at (0.7,4) {24};
        \node (25) at (2.2,4) {25};
        \node (34) at (3.7,4) {34};
        \node (35) at (5.2,4) {35};
        \node (45) at (6.7,4) {45};
        % Level 1
        \node (1) at (-4,2) {1};
        \node (2) at (-2,2) {2};
        \node (3) at (0,2) {3};
        \node (4) at (2,2) {4};
        \node (5) at (4,2) {5};
        % Level 0
        \node (0) at (0,0) {$\emptyset$};
        
        % Connections
        \foreach \i in {1,2,3,4,5}
            \draw[blue!90!white,dashed,thick] (0) -- (\i);
        \draw[black!20!white,dashed] (1) -- (12)
            (1) -- (13)
            (1) -- (14)
            (1) -- (15)
            (2) -- (12);
        \draw[blue!90!white,dashed,thick] (2) -- (23)
             (2) -- (12)
            (2) -- (24)
            (2) -- (25);
        \draw[black!20!white,dashed] (3) -- (13)
            (3) -- (23)
            (3) -- (34)
            (3) -- (35)
            (4) -- (14)
            (4) -- (24)
            (4) -- (34)
            (4) -- (45)
            (5) -- (15)
            (5) -- (25)
            (5) -- (35)
            (5) -- (45)
            (12) -- (123)
            (12) -- (124)
            (12) -- (125);
        \draw[black!20!white,dashed] (13) -- (123)
            (13) -- (134)
            (13) -- (135)
            (14) -- (124)
            (14) -- (134)
            (14) -- (145)
            (23) -- (234)
            (23) -- (123)
            (23) -- (235)
            (24) -- (234)
            (24) -- (124)
            (24) -- (245)
            (34) -- (134)
            (34) -- (234)
            (34) -- (345)
            (15) -- (125)
            (15) -- (145)
            (15) -- (135)
            (25) -- (245)
            (25) -- (235)
            (25) -- (125)
            (35) -- (135)
            (35) -- (235)
            (35) -- (345)
            (45) -- (145)
            (45) -- (245)
            (45) -- (345);
        \draw[black!20!white,dashed] (123) -- (1234)
            (123) -- (1235)
            (125) -- (1235)
            (125) -- (1245);
        \draw[black!20!white,dashed] (134) -- (1234)
            (134) -- (1345)
            (234) -- (1234)
            (234) -- (2345)
            (124) -- (1234)
            (124) -- (1245)
            (135) -- (1345)
            (135) -- (1235)
            (145) -- (1245)
            (145) -- (1345)
            (235) -- (1235)
            (235) -- (2345)
            (245) -- (1245)
            (245) -- (2345)
            (345) -- (2345)
            (345) -- (1345);
        \draw[black!20!white,dashed] (1234) -- (12345)
            (1235) -- (12345)
            (1345) -- (12345)
            (1245) -- (12345)
            (2345) -- (12345);
        \draw[blue!90!white,dashed,thick] (25) -- (245)
        (25) -- (125)
        (23) -- (123)
        (23) -- (234)
        (123) -- (1234)
        (123) -- (1235);
        \draw[->,blue!50!black,very thick] (0) -- (2);
        \draw[->,blue!50!black,very thick] (2) -- (25);
        \draw[->,blue!50!black,very thick] (25) -- (235);
        \draw[->,blue!50!black,very thick] (23) -- (123);
        \draw[->,blue!50!black,very thick] (123) -- (1234);
        \draw[->,red!50!black,very thick] (235) -- (23);
        \end{tikzpicture}
    \label{fig::rewiring}    
    \text{(c) \textsf{ResQue Greedy} Algorithm.}
    \end{minipage}
    \caption{Hasse diagram for a ground set of five elements $\mathcal{P}=\{1,2,3,4,5\}$. In a), the paths created by the sequential greedy algorithm (arrows) and the observed solutions of the sequential greedy algorithm (dashed). In b), all the possible paths that lead to the optimal solution $\mathcal{S}^\star_4=\{1,2,3,4\}$ and in c), an example of the rewiring operation over the lattice where blue lines stand for the paths created by the sequential greedy algorithm (arrows) and the observed solutions of the sequential greedy algorithm (dashed) and the red lines are the edge added by the rewiring operation at stage $3$. }
    \label{fig::lattice_combined2}
\end{figure*}
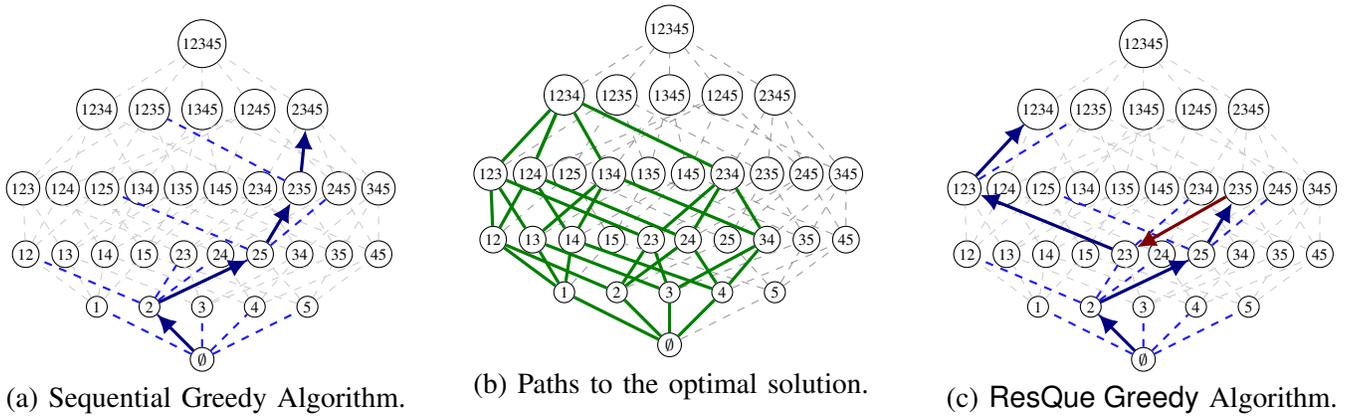
In addition to total curvature, other notions of curvature have also been proposed in the literature. One notable example is when curvature is defined with respect to the optimal set. With such curvature $c^\star$, the bound $\frac{1}{c^\star}(1-\text{e}^{-c^\star})$ has been proven to be the tightest possible in polynomial time. Specifically, \cite{JV:10} has demonstrated that achieving any tighter approximation  would require an exponential number of value queries. The interested reader can explore other literature on the role of curvature in improving optimality bounds in other work such as \cite{BL-BVO-EC-AP:24}, where they reformulate a new curvature definition for string submodular functions such that it is computationally achievable, or in \cite{SG-HH-AK:15} they define a linear curvature for Probabilistic Submodular Models.
On the other hand, there is extended literature on how to adapt curvature notion to many particular sub-problems on the field. For example, in \cite{HH-MS-AK:17} they use curvature notion to define weak-submodularity on functions and introduce submodular online optimization.

\emph{Statement of contribution}: 
Improving optimality gap of suboptimal solution for resource allocation problems is of paramount importance as it can lead to substantial financial savings and efficient utilization of resources; enhancing allocation strategies in various domains, such as logistics, supply chain management, sensor scheduling, or robotic coverage. Such improvement can considerably reduce costs and improve service levels. Motivated by this impact, this paper proposes Rewired Sequential Greedy (\textsf{ResQue Greedy}), a novel framework that dynamically adjusts the solution path of the standard sequential greedy algorithm to improve the  final polynomial time solution. Specifically, \textsf{ResQue Greedy} mitigates the sequential greedy algorithm's inherent shortsightedness by assessing the impact of current choices on future gains, leveraging a localized, polynomial-time computable curvature metric. If an element in the sequential solution is deemed harmful to future marginal gains, \textsf{ResQue Greedy} eliminates it, allowing the greedy algorithm to select a potentially better substitute. We demonstrate that \textsf{ResQue Greedy}'s worst case optimality gap is $1-\text{e}^{-1}$, but in practice, it achieves better performance compared to the standard sequential greedy algorithm. Unlike existing literature~\cite{BS-SLS:17,JV-SSK:24} that examine the impact of solution path disruptions, our work explores actively modifying the sequential greedy path for improved gain. This approach refines the traditional greedy strategy and establishes a foundation for curvature-aware methodologies, yielding better approximation guarantees with minimal computational overhead. By focusing on practical improvements in optimality bounds, \textsf{ResQue Greedy} paves the way for a new class of optimization algorithms with significant potential for advancing submodular maximization under cardinality constraints.

%%%%%%%%%%%%%%%%%%%%%%%%%%%%%%%%%%%%%%%%%%%%%%
%%%%%%%%%%%%%%%%%%%%%%%%%%%%%%%%%%%%%%%%%%%%%%
%%%%%%%%%%%%%%%%%%%%%%%%%%%%%%%%%%%%%%%%%%%%%%
\section{Motivation and Objective statement}
The sequential greedy algorithm, Algorithm~\ref{alg:sequential}, is a widely utilized method for solving Problem~\eqref{eq::mainProblem}, offering a theoretical guarantee of near-optimal solutions with a polynomial number of function evaluations. This is particularly advantageous compared to exhaustive search, which exhibits exponential complexity with respect to the ground set's cardinality. The central objective of this paper is to understand the sub-optimality of the greedy algorithm, despite its iterative selection of elements with maximal marginal gain, i.e., $\Delta_f(x_{i+1}|\mathcal{S}_i) \geq \Delta_f(x_j|\mathcal{S}_i),~\forall~x_j\in\mathcal{P}\setminus\mathcal{S}_i$ and seek modifications to the algorithm that would lead to a better optimality gap. 

To start our study, we conceptualize any sequential solution construction process for Problem~\eqref{eq::mainProblem} as a path traversal on the lattice representation of the ground set's power set, visualized by the Hasse diagram (Fig.~\ref{fig::lattice_combined2}). The sequential greedy algorithm can be interpreted as a step-wise path-finding procedure, starting from the empty set and progressing through the lattice until the final solution is obtained. In Fig.~\ref{fig::lattice_combined2} (a), the highlighted blue path, leading to set $\mathcal{S}=\{2,3,4,5\}$, exemplifies this process. A defining characteristic of this path is the algorithm's restricted evaluation at each stage $i$, where only edges originating from $\mathcal{S}_{i-1}$, shown by dashed lines in Fig.~\ref{fig::lattice_combined2} (a), are considered, thus confining the search to a limited subset of possible transitions.

For the sake of demonstration, let the optimal solution be $\mathcal{S}^\star=\{1,2,3,4\}$ as shown in Fig.~\ref{fig::lattice_combined2} (b). As we can see in the figure, there are multiple paths highlighted in green over the Hasse diagram that lead to the eventual optimal solution. Sequential greedy algorithm, similar to any greedy algorithm suffers from shortsightedness, that is, the current decisions can substantially impact the outcome going forward. This is clearly illustrated in Fig.~\ref{fig::lattice_combined2} (a) where the course of sequential greedy is deviated from the optimal path towards  $\mathcal{S}^\star=\{1,2,3,4\}$ due to the decision made in stage $2$, going from $\mathcal{S}=\{2\}$ to $\mathcal{S}=\{2,5\}$. Notice that going forward from $\mathcal{S}=\{2,5\}$, recovering the optimal solution is impossible as the sequential greedy algorithm will never consider the edges belonging to an optimal path, shown in green Fig.~\ref{fig::lattice_combined2} (b).
This visualization suggests that the sequential process could potentially benefit from strategic diversions or ``rewiring" to redirected towards the optimal solution. This observation raises the fundamental question: how can such rewiring be effectively conducted?

One intuitive approach involves, at each stage $i$, comparing the function value at the greedy selection $\mathcal{S}_i$ with randomly generated sets $\mathcal{S}'_1,\cdots,\mathcal{S}'_m$, where $\mathcal{S}'_j\cap\mathcal{S}_{i}=\emptyset$ and $|\mathcal{S}'_j|=i$, $j\in\{1,\cdots,m\}$. Evaluating the function value for these sets and reinitializing the path with a high-value set could potentially improve the solution. However, this incurs substantial computational overhead. Therefore, the core research question addressed in this paper is: How can we design a trigger mechanism to dynamically redirect the sequential greedy algorithm's path on the Hasse diagram when necessary, achieving improved approximation guarantees with reasonable computational cost?

Given the established $(1-\text{e}^{-1})$ approximation barrier for polynomial-time algorithms, our objective is not to surpass this worst case bound. Instead, we aim to leverage the function's curvature to derive tighter approximation guarantees that reflect the specific submodular characteristics of the problem, thereby achieving improvements over the baseline $\frac{1}{c}(1-\text{e}^{-c})$ bound, where $c$ is the total curvature.

%%%%%%%%%%%%%%%%%%%%%%%%%%%%%%%%%
%%%%%%%%%%%%%%%%%%%%%%%%%%%%%%%%%%
%%%%%%%%%%%%%%%%%%%%%%%%%%%%%%%%%%%%%%%%%%%%%%%%%%%%%%%%%%%%%%%%%%%%%%%%%%%%%%%%%%%%%%%%%%%%%%%%

\section{Set Curvature}
\label{sec::local_curvature}
The marginal gain, $f(\mathcal{S} \cup \{e\}) - f(\mathcal{S})$, for $e \in \mathcal{A}$, serves as the primary guide for the sequential greedy algorithm, indicating the immediate value of adding each element $e$. To address the shortsightedness in this process and enable dynamic path correction, we propose a predictive mechanism based on the notion of curvature. 

The total curvature $c$ represents the worst case diminishing returns, but this worst case scenario  may not be relevant to the solution of problem~\eqref{eq::mainProblem} (e.g., $c=1$ due to a specific subset of $\mathcal{P}$, which may not be in feasible set of~\eqref{eq::mainProblem}). Therefore, we shift our focus to local curvature notions, specifically those tied to sequential solution building. We begin by defining \emph{set curvature}, which quantifies the impact of a set $\mathcal{S}$ on the potential value extractable from permissible expansion elements  in $\mathcal{A} \subset \mathcal{P}\setminus\mathcal{S}$.

\begin{defn}[Set curvature]
    Let $f:2^\mathcal{P}\to\real_{\geq0}$ be monotone increasing normal submodular function over $\mathcal{P}$. The \emph{set curvature} of set $\mathcal{S}\subset\mathcal{P}$ with respect to the permissible expansion $\mathcal{A}\subset\mathcal{P}\setminus\mathcal{S}$ is defined as:
    \begin{align}\label{eq::set_curve}
    \gamma(\mathcal{S}|\mathcal{A}) &=
    1 - \min_{\substack{e \in \mathcal{A}}} \frac{f(\mathcal{S} \cup \{e\}) - f(\mathcal{S})}{f(\{e\}) - f(\emptyset)}.
\end{align}
\boxend
\end{defn}
Notice that for a set curvature, because of $f:2^\mathcal{P}\to\real_{\geq0}$ being normal and submodular, we have $f(\mathcal{S} \cup \{e\}) - f(\mathcal{S})\leq f(e)$. Therefore, $\gamma(\mathcal{S}|\mathcal{A})\leq 1$. 
Moreover, the next result shows that set curvature is monotone increasing with respect to the same permissible expansion.
\begin{lem}\longthmtitle{Set curvature is monotone increasing}\label{lem::monotonicity}
    Let $f:2^\mathcal{P}\to\mathbb{R}^{+}$ be a monotone increasing and normal submodular function over the ground set $\mathcal{P}$. Given $\mathcal{S}_1\subset\mathcal{S}_2\subset\mathcal{P}$, and any $\mathcal{A}\subset\mathcal{P}\setminus\mathcal{S}_2$, we have $\gamma(\mathcal{S}_1|\mathcal{A})\leq \gamma( \mathcal{S}_2|\mathcal{A}).$ 
\end{lem}
\begin{proof}
  The proof follows from taking into account the submodularity of $f$ when computing the set curvature~\eqref{eq::set_curve}. Submodularity of $f$ results in $f(\mathcal{S}_2 \cup \{e\}) - f(\mathcal{S}_2)\leq f(\mathcal{S}_1 \cup \{e\}) - f(\mathcal{S}_1)$ for any $e\in\mathcal{A}$.
\end{proof}

Note that given $\gamma(\mathcal{S}|\mathcal{A})$, we can write $f(\mathcal{S} \cup \{e\}) - f(\mathcal{S})\geq (1-\gamma(\mathcal{S}|\mathcal{A}))f(e)$ for all $e\in\mathcal{A}$. Consequently, if $\gamma(\mathcal{S}_1|\mathcal{A}) \leq \gamma(\mathcal{S}_2|\mathcal{A})$ for two sets $\mathcal{S}_1$ and $\mathcal{S}_2$, we can infer that $\mathcal{S}_1$ is more likely to yield a better expansion value compared to $\mathcal{S}_2$. This predictive capability of the set curvature allows us to anticipate potential impact of added elements during the sequential expansion of $\mathcal{S}$ to solve~\eqref{eq::mainProblem}, providing a foundation for our dynamic rewiring mechanism. To pave the way towards this mechanism, next, for any sequential solution built from $\mathcal{P}$, we define the \emph{expansion curvature} and the \emph{path curvature} as follows. 

\begin{defn}[Expansion and path curvatures]\label{def::expansion_curvatue_path}
Let $f:2^\mathcal{P}\to\real_{\geq0}$ be monotone increasing normal submodular function over $\mathcal{P}$. Consider a sequential set building where  $\mathcal{S}_0=\emptyset$ and  $\mathcal{S}_i=\mathcal{S}_{i-1}\cup \{e\}$ where $e\in\mathcal{A}_{i-1}=\mathcal{P}\backslash\mathcal{S}_{i-1}$ for $i\in\{1,\cdots,\kappa\}$, $\kappa \leq |\mathcal{P}|$. We define the \emph{expansion curvature} of this sequential solution building up to stage $i\in\{1,\cdots,\kappa\}$ as 
\begin{equation}
    \label{eqn::greedy_curvature}
    \gamma_{\text{e}}(\mathcal{S}_i) = \max\left\{ \gamma(\mathcal{S}_0|\mathcal{A}_0),\cdots,\gamma(\mathcal{S}_{i-1}|\mathcal{A}_{i-1}) \right\}.
\end{equation}
We define the \emph{path curvature} of this sequential solution building up to stage $i$ as 
    \begin{equation}
        \label{eqn::greedy_curvature2}
        \gamma_p(\mathcal{S}_i) = \max\left\{ \gamma(\mathcal{S}_0|s_1) ,\gamma(\mathcal{S}_1|s_2) \cdots, \gamma(\mathcal{S}_{i-1}|s_{i}) \right\}.
    \end{equation}
    \boxend
\end{defn}
The result below highlight some of the properties of the expansion and path curvatures.
\begin{lem}[Properties\! of\! expansion\! and\! path\! curvatures]
    \label{lem::properties}
    Consider the expansion and path curvatures defined in Definition~\ref{def::expansion_curvatue_path}. Then, for any $i\in\{1,\cdots,\kappa\}$ we have
    \begin{itemize}
        \item [(a)] $\gamma_{\text{e}}(\mathcal{S}_\ell)\leq \gamma_{\text{e}}(\mathcal{S}_i)$, and $\gamma_{\text{p}}(\mathcal{S}_\ell)\leq \gamma_{\text{p}}(\mathcal{S}_i),$ for all $\ell\leq i. $
        \item [(b)] $\gamma_{\text{e}}(\mathcal{S}_i)=\max\{\gamma_p(\mathcal{S}_{i}),\gamma(\mathcal{S}_{i-1}|\mathcal{A}_{i-1})\}=\max\{\gamma_p(\mathcal{S}_{i-1}),\gamma(\mathcal{S}_{i-1}|\mathcal{A}_{i-1})\}.$
        \item [(c)] $\gamma_{\text{p}}(\mathcal{S}_i)\leq \gamma_{\text{e}}(\mathcal{S}_i)$.
    \end{itemize}
\end{lem}
\begin{proof}
Part (a) follows trivially from the definition of the expansion curvature at any stage. The proof of part (b) follows from definition of set curvature that ensures $\gamma(\mathcal{S}_\ell|\mathcal{A}_\ell)=\gamma(\mathcal{S}_\ell|\mathcal{A}_{\ell+1}\cup\{s_{\ell+1}\})=\max\{\gamma(\mathcal{S}_\ell|\mathcal{A}_{\ell+1}),\\\gamma(\mathcal{S}_\ell|\{\ell_{\ell+1}\})\}$ and invoking Lemma~\ref{lem::monotonicity} which indicates $\gamma(\mathcal{S}_\ell|\mathcal{A}_{\ell+1})\leq \gamma(\mathcal{S}_{\ell+1}|\mathcal{A}_{\ell+1})$ because $\mathcal{S}_\ell\subset \mathcal{S}_{\ell+1}$. Part (c) follows from part (b).
\end{proof}

Another set curvature that will be relevant in our rewiring algorithm design is 
the \emph{path curvature} of a sequential solution with respect to $\mathcal{S}^\star$, a maximizer of~\eqref{eq::mainProblem}. Specifically, based on \eqref{eqn::greedy_curvature2}, let 
\begin{equation}
\label{eq::gamma_star}
    \gamma^\star=\max\{\gamma(\mathcal{S}^\star\cup\mathcal{S}_{0}|s_1),\cdots,\gamma(\mathcal{S}^\star\cup\mathcal{S}_{\kappa-1}|s_\kappa)\}
\end{equation}
This curvature, which is equivalent to $c^\star$ in Conforti et al.~\cite{MC-GC:84}, gives the tightest optimality bound   $\frac{1}{c^\star}(1-\text{e}^{-c^\star})$ achievable in polynomial time for~\eqref{eq::mainProblem}. Invoking Lemma~\ref{lem::monotonicity}, from definitions of $\gamma^\star$ and $\gamma_{\text{p}}$ we can conclude that
\begin{equation}
    \label{eq::curvature_relation}
         \gamma^\star\geq \max\gamma_{\text{p}}(\mathcal{S}_i),\quad i\in\{1,\cdots,\kappa\}.
\end{equation}

\begin{algorithm}[t]
{\small
\caption{High level description of \textsf{ResQue Greedy}}\label{alg:resque_high}
\begin{algorithmic}[1]
\Require Ground set $\mathcal{P}$ and utility function $f$
\Ensure $\mathcal{S}^{\text{RSG}}\subset \mathcal{P}$  satisfying $|\mathcal{S}^{\text{RSG}}|\leq \kappa$
\State $\mathcal{S}_0 \gets \emptyset$
  \For{$i\in\{0,\cdots\kappa-1\}$}
  \State $s_{i+1} \gets \text{argmax}_{s \in \mathcal{P} \backslash \mathcal{S}_{i}}{ \Delta_f(s|\mathcal{S}_{i})}$
  \State $\bar{\mathcal{S}}_{i+1}\gets \mathcal{S}_{i}\cup\{s_{i+1}\}$
  \If{Trigger law is activated}
  \State $\mathcal{S}^\prime_{i}=\bar{\mathcal{S}}_{i+1}\backslash\{\bar{s}\}$ where $\bar{s}\in\bar{\mathcal{S}}_{i+1}$ is the element removed base on the step-back policy
  \State $s_{i+1} \gets \text{argmax}_{s \in \mathcal{P} \backslash \mathcal{S}^\prime_i}{ \Delta_f(s|\mathcal{S}^\prime_i)}$
   \State $\mathcal{S}_{i+1}\gets \mathcal{S}'_{i}\cup\{s_{i+1}\}$
  \Else
   \State $\mathcal{S}_{i+1}\gets \bar{\mathcal{S}}_{i+1}$
  \EndIf
  \EndFor
  \State $\mathcal{S}^{\text{RSG}}\gets\mathcal{S}_\kappa$
\end{algorithmic}
}
\end{algorithm}

\section{\textsf{ResQue Greedy}: Rewiring Sequential Greedy Algorithm to Improve optimality Gap}
\label{sec::resque}
Algorithm~\ref{alg:resque_high}, presents our proposed Rewired Sequential Greedy (\textsf{ResQue Greedy}) Algorithm. \textsf{ResQue Greedy} is a modified variation of the standard sequential greedy approach, Algorithm~\ref{alg:sequential}, that incorporates a rewiring mechanism. This allows for dynamic adjustments to the ``solution path". For example, as illustrated in Fig.~\ref{fig::lattice_combined2}(c), the algorithm initially follows the sequential greedy trajectory, constructing sets $\mathcal{S}_1=\{2\}$, $\mathcal{S}_2=\{2,5\}$ and $\mathcal{S}_3=\{2,3,5\}$. Upon triggering a predefined rewiring condition at stage $3$, the algorithm does an step-back, removing an element, in this case $5$, based on a rewiring step-back criterion to obtain $\mathcal{S}^\prime_2=\{2,3\}$. Subsequently, it explores an alternative path based on greedy selection of the element yielding highest marginal gain to arrive at rewired $\mathcal{S}_3=\{1,2,3\}$ and continuing with sequential greedy selection ultimately reaching the solution $\mathcal{S}^\star=\{1,2,3,4\}$. Reaching to $\mathcal{S}^\star$ is not guaranteed but this case demonstrated the possibility that rewiring offers. 
%Notably, all steps, except the rewiring process, adhere to the standard sequential greedy policy of selecting the element with the maximum marginal gain. 
It is important to note that after rewiring, the sequential greedy process, based on adding the element with the highest marginal gain, can potentially recover the original set $\mathcal{S}_3=\{2,3,5\}$ if it offers a higher marginal gain than the newly introduced possibilities due to the rewiring. Due to space limitation, the illustration in Fig.~\ref{fig::lattice_combined2} shows a small scale problem and a single rewiring step. The \textsf{ResQue Greedy} may implement multiple rewiring along its path.

The following result states that regardless of the trigger law and the element elimination policy for rewiring, \textsf{ResQue Greedy} 
is guaranteed to yield an optimality gap of no worse than $(1-\text{e}^{-1})$.

\begin{thm}[Optimality gap of \textsf{ResQue Greedy}]
    Let $f:2^\mathcal{P}\to\mathbb{R}^{+}$ be a monotone increasing and normal submodular function over the ground set $\mathcal{P}$. 
Consider \textsf{ResQue Greedy} given in Algorithm~\ref{alg:resque_high} under the assumption that the trigger law never activates at step $1$\footnote{Note that rewiring is irrelevant at step $i=1$, because it will lead to repetition of the same greedy selection process.}. Then,
        \begin{equation}\label{eq::greedy_step}
        f(\mathcal{S}_{i})\geq\left( 1-(1-\frac{1}{\kappa})^i\right)f(\mathcal{S}^\star),
    \end{equation}
    for $i\in\{1,\cdots,\kappa\}$, which leads to
    $f(\mathcal{S}^\text{RSG})\geq (1-\text{e}^{-1})f(\mathcal{S}^\star).$ 
    \end{thm}
    \smallskip
\begin{proof}
    This proof is done by induction. To confirm~\eqref{eq::greedy_step} holds at $i=1$ we proceed as follows. Starting at $\mathcal{S}_0=\emptyset$, due to greedy selection to build $\mathcal{S}_1$, we know that $f(\mathcal{S}_1)\geq \frac{1}{\kappa}\sum_{i=1}^\kappa f(s_i^\star)$. Due to submodularity of $f$, we have  $\sum_{i=1}^\kappa f(s_i^\star)\geq f(\mathcal{S}^\star)$. Therefore, $f(\mathcal{S}_1)\geq \frac{1}{\kappa}f(\mathcal{S}^\star)$, which confirms~\eqref{eq::greedy_step} holds at the first step of the algorithm. 

Next let~\eqref{eq::greedy_step} holds up to step $i$. We will show~\eqref{eq::greedy_step} holds at step $i+1$ as well. If step $i+1$ is the result of a greedy selection without rewiring (trigger law is not activated), $\mathcal{S}_{i+1}$ is given by line 10 of Algorithm~\ref{alg:resque_high} where $\bar{\mathcal{S}}_{i+1}$ is computed at line 4. In this case we can write
 \begin{equation}\label{eq::uniform_bound_proof}
    \left.\begin{array}{lll}
        f(\mathcal{S}^\star)&\leq f(\mathcal{S}^\star\cup\mathcal{S}_i)&\text{(monotone increasing)}\\
        &=f(\mathcal{S}_i)+\sum_{j=1}^\kappa \nolimits\Delta_f(s_j^\star|\mathcal{S}_i\cup\{s_1^\star,\cdots,s_{j-1}^\star\})&\text{(telescoping sum)}\\&\leq f(\mathcal{S}_i)+\sum_{j=1}^\kappa \nolimits\Delta_f(s_j^\star|\mathcal{S}_i)&\text{(submodularity)}\\
        &\leq f(\mathcal{S}_i)+\sum\nolimits\nolimits_{j=1}^\kappa(f(\mathcal{S}_{i+1})-f(\mathcal{S}_i))&\text{(greedy selection to build }\mathcal{S}_{i+1}=\bar{\mathcal{S}}_{i+1})\\
        &=f(\mathcal{S}_i)+\kappa \,(f(\mathcal{S}_{i+1})-f(\mathcal{S}_i)),
        \end{array}\right.
    \end{equation}
    which leads to
%$$\frac{1}{\kappa}f(\mathcal{S}^\star)\leq -(1-\frac{1}{\kappa})f(\mathcal{S}_i)+f(\mathcal{S}_{i+1})$$
$f(\mathcal{S}_{i+1})\geq \frac{1}{\kappa}f(\mathcal{S}^\star)+(1-\frac{1}{\kappa})f(\mathcal{S}_i).$
    Subsequently, given that~\eqref{eq::greedy_step} holds at step $i$, we can conclude that 
    $$\!f(\bar{\mathcal{S}}_{i+1})\!=\!f(\mathcal{S}_{i+1})\!\geq\! \frac{1}{\kappa}f(\mathcal{S}^\star)+(1-\frac{1}{\kappa})\!\left(\! 1-(1\!-\frac{1}{\kappa})^i\!\right)\!f(\mathcal{S}^\star),$$
    confirming that~\eqref{eq::greedy_step} holds at $i+1$.

Now consider the case that the rewiring trigger law is activated %From now on, for the sake of understanding, we will refer to the set after a rewiring operation as $\mathcal{S}^\prime$. 
and the element $\bar{s}$ of $\bar{\mathcal{S}}_{i+1}$ computed at line 4 of Algorithm~\ref{alg:resque_high} is removed to lead to $\mathcal{S}^\prime_{i}$ given in line 6. For $\bar{\mathcal{S}}_{i+1}$ before step-back and rewiring, we have already shown that 
        \begin{equation}\label{eq::S_i+1_before_rewire}
        f(\bar{\mathcal{S}}_{i+1})\geq\left( 1-(1-\frac{1}{\kappa})^{i+1}\right)f(\mathcal{S}^\star).
    \end{equation}
By virtue of the greedy expansion from $\mathcal{S}'_i$ and the fact that $\bar{\mathcal{S}}_{i+1}=\mathcal{S}'_i\cup\{\bar{s}\}$ after step-back and executing lines 7 and 8 of Algorithm~\ref{alg:resque_high}, we have $f(\mathcal{S}_{i+1})\geq f(\bar{\mathcal{S}}_{i+1})$, which together with~\eqref{eq::S_i+1_before_rewire} confirm that~\eqref{eq::greedy_step} holds in the rewiring scenario, as well. This completes our induction proof argument. Since we confirmed~\eqref{eq::greedy_step} holds for every $i\in\{1,\cdots,\kappa\}$, following using\footnote{This argument is similar to the one used in proof of the conventional sequential greedy gap; see~\cite{SSK:25} for details.} $(1-\frac{1}{\kappa})^\kappa\leq \text{e}^{-1}$ for any $\kappa> 1$, we can confirm that for $\mathcal{S}^\text{RSG}=\mathcal{S}_\kappa$ given in line 13 of Algorithm~\ref{alg:resque_high}, we have  $f(\mathcal{S}^\text{RSG})\geq (1-\text{e}^{-1})f(\mathcal{S}^\star)$.
\end{proof}

\medskip
%This means that the \textsf{ResQue Greedy} algorithm regardless of the trigger law is guaranteed to yield an optimality gap of no worse than $1-\frac{1}{e}$. 
Despite similar worst case optimality gap, \textsf{ResQue Greedy} is  expected to achieve a tighter actual gap compared to the standard sequential greedy algorithm due to $f(\mathcal{S}_{i+1})\geq f(\bar{\mathcal{S}}_{i+1})$ after rewiring (lines 6 to 8 of Algorithm~\ref{alg:resque_high}). 

Observe that if we perform a greedy selection to construct $\mathcal{S}_{i+2}$ and $\bar{\mathcal{S}}_{i+2}$, respectively from $\mathcal{S}_{i+1}$ computed at line 8 and $\bar{\mathcal{S}}_{i+1}$ computed at line 4 of Algorithm~\ref{alg:resque_high}, we have 
\begin{subequations}\label{eq::expansion_bound}
    \begin{align}
        f(\mathcal{S}_{i+2})\geq \frac{1}{\kappa}f(\mathcal{S}^\star)+(1-\frac{1}{\kappa})f(\mathcal{S}_{i+1}),\\
        f(\bar{\mathcal{S}}_{i+2})\geq \frac{1}{\kappa}f(\mathcal{S}^\star)+(1-\frac{1}{\kappa})f(\bar{\mathcal{S}}_{i+1}).
    \end{align}
\end{subequations}
Since $f(\mathcal{S}_{i+1}) \geq f(\bar{\mathcal{S}}_{i+1})$, it follows that in equation~\eqref{eq::expansion_bound}, $f(\mathcal{S}_{i+2})$ possesses a larger lower bound and is \emph{likely} to achieve a higher value compared to $f(\bar{\mathcal{S}}_{i+2})$. A straightforward scenario where \textsf{ResQue Greedy} guarantees superior performance over standard sequential greedy is when rewiring occurs at the final step. However, due to the diminishing returns property of submodular functions, the potential gains from rewiring at this stage may be limited. Nonetheless, regardless of when rewiring is performed, a critical question remains: which element's removal is more likely to result in a strict increase in value for the subsequent step, specifically, $f(\mathcal{S}_{i+1}) > f(\bar{\mathcal{S}}_{i+1})$?

Moreover, the potential gain in optimality through rewiring incurs increased computational overhead. Specifically, each rewiring step at stage $i+1$ necessitates a step-back, followed by an additional greedy iteration with $n - i$ queries. Consequently, frequent rewiring can lead to an exponential increase in computational complexity. To mitigate this computational burden while preserving the benefits of rewiring, we propose a trigger law based on the expansion curvature introduced in Section~\ref{sec::local_curvature}. Recognizing that precise predictions of future marginal gains are often unattainable, this trigger law determines when rewiring is likely to yield a significant improvement in optimality, thereby limiting unnecessary rewiring steps. By effectively managing rewiring, we aim to control computational complexity while enhancing the likelihood that \textsf{ResQue Greedy} outperforms the standard sequential greedy algorithm.

%Moreover, the potential gain in optimality through rewiring incurs increased computational overhead. Specifically, each rewiring step at stage $i+1$ necessitates a step-back, followed by an additional greedy iteration with $n - i$ queries. Consequently, frequent rewiring can lead to an exponential increase in computational complexity. To mitigate this computational burden while preserving the benefits of rewiring, we propose a trigger law based on the expansion curvature introduced in Section~\ref{sec::local_curvature} to execute rewiring when it is deemed more effective. Recognizing that absolute guarantees of future behavior are unattainable, we employ. This approach allows us to control complexity while simultaneously enhancing the likelihood that \textsf{ResQue Greedy} outperforms the standard sequential greedy algorithm.

%%%%%%%%%%%%%%%%%%%%%%%%%%%%%%%%%%%%%%

\subsection{Trigger Law Design}
To define our trigger law, we turn to the concept of curvature and its predictive nature in quantifying the expected return of set expansion  and the role of set curvature in qualifying the optimality gap discussed in Section~\ref{sec::local_curvature}. Our design is motivated by two key observations.

\begin{comment}
\begin{algorithm}[t]
{\small
\caption{Trigger law and step back policy}\label{alg:fault}
\begin{algorithmic}[1]
    \Require Constructed set $\bar{\mathcal{S}}_{i+1}$, expansion curvature $\gamma_e(\mathcal{S}_{i})$ and set curvature $\gamma(\mathcal{S}_{i}|\mathcal{A}_i)$
    \Ensure Updated $\mathcal{S}^\prime_i$, $\gamma_e(\mathcal{S}^\prime_{i})$
    \If{$\gamma(\mathcal{S}_{i}|\mathcal{A}_{i}) \leq \gamma_e(\mathcal{S}_{i})$ $\longleftarrow$ Trigger law
        \State Identify $\bar{s}=s_{j_i}$ [Definition~\ref{defn::policy}]
        \State Update set $\mathcal{S}^\prime_i \leftarrow \bar{\mathcal{S}}_{i+1} \setminus \{\bar{s}\}$
        \For{$\ell\in\{1,\cdots,i+1\}$}
        \If{$\ell = j$}
        \State Delete $\gamma(\mathcal{S}_\ell|\mathcal{A}_\ell)$
        \ElsIf{$\ell > j$}
        \State Update $\gamma(\mathcal{S}^\prime_\ell|\mathcal{A}^\prime_\ell)$ as \eqref{eqn::update_rule}
        \EndIf
        \EndFor
        \State Recompute $\gamma_e(\mathcal{S}^\prime_i) \!\gets\! \max \left\{ \gamma(\mathcal{S}^\prime_{\ell-1}|\mathcal{A}_{\ell-1}) \right\}_{\ell=1,\cdots,i}$
    \EndIf
\end{algorithmic}
}
\end{algorithm}
\end{comment}
The first insight comes from the conceptual motivation of the \textsf{ResQue Greedy} as a redirecting method over the sequential solution building path on the Hasse diagram of the ground set to an ``optimal solution path" as illustrated in Fig.~\ref{fig::lattice_combined2}. Let us consider $\Delta_f(s_i|\mathcal{S}^\star\cup\mathcal{S}_{i-1})$, which is zero if $s_i\in\mathcal{S}^\star$. Otherwise, if $s_i\notin\mathcal{S}^\star$, given the definition of $\gamma^\star$ in~\eqref{eq::gamma_star}, and submodularity of $f$, we have 
$$f(s_i)\geq \Delta_f(s_i|\mathcal{S}^\star\cup\mathcal{S}_{i-1})\geq (1-\gamma^\star)f(s_i).$$
To achieve maximal gain of adding $s_i$ we want $\gamma^\star$ to be small.  This is aligned with expectation for small $\gamma^\star$ that can be deduced from the best optimality gap $\frac{1}{\gamma^\star}(1-\text{e}^{-\gamma^\star})$. From, \eqref{eq::curvature_relation} and $\gamma_\text{p}$'s monotone increasing nature, we can see that smaller path curvature $\gamma_{\text{p}}$ will increase the chance of smaller $\gamma^\star$. Therefore, in rewiring process, we should seek for a small \emph{path curvature} $\gamma_p(\mathcal{S}_i)$. 
The value of keeping $\gamma_p(\mathcal{S}_i)$ small shows itself in the following relations
\begin{align*}
   f(\mathcal{S}_i)-f(\mathcal{S}_{i-1}) &= (1-\gamma(\mathcal{S}_{i-1}|s_{i}))f(s_i),\\
   &~~\vdots\\
f(\mathcal{S}_1)-f(\mathcal{S}_{0}) &= (1-\gamma(\mathcal{S}_0|s_{1}))f(s_1), 
\end{align*}
leading to $f(\mathcal{S}_i)\geq \left(1-\gamma_p(\mathcal{S}_i)\right)\sum_{s\in\mathcal{S}_i}f(s)$. This indicates again that we desire small $\gamma_p(\mathcal{S}_i)$ for higher value of $f(\mathcal{S}_i)$.

The second observation we make is based on how $\gamma_{\text{p}}$ evolves along the path of sequential solution building. Invoking Lemma~\ref{lem::properties}, we can anticipate the rate of growth of $\gamma_p(\mathcal{S}_{i})$  by knowing that $\gamma_p(\mathcal{S}_{i})$ is upper bounded by $\gamma_e(\mathcal{S}_{i})$. By the monotonicity in \emph{expansion curvature}, $\gamma_e(\mathcal{S}_{i})\leq\gamma_e(\mathcal{S}_{i+1})$. Therefore, the greater the expansion curvature  $\gamma_e(\mathcal{S}_{i})$, the greater the path curvature $\gamma_p(\mathcal{S}_{i+1})$ is expected in the worst case scenario.

Given the aforementioned observations, to control the expected worst-case path curvature, our trigger law monitors the expansion curvature as described below.
\begin{comment}
Leveraging the relationship between set and expansion curvature, the trigger law determines when rewiring is beneficial. Recall that set curvature $\gamma(\mathcal{S}_{i-1}|\mathcal{A}_{i-1})$ measures the diminishing returns at iteration $i$, while expansion curvature $\gamma_{\text{e}}(\mathcal{S}_{i-1})$ represents the maximum set curvature observed up to that point. Intuitively, a gradual increase in expansion curvature indicates a smooth decrease in permissible expansion, suggesting no single element $s\in\mathcal{S}_{i}$ significantly dominates future diminishing returns. Conversely, a sudden drop in permissible expansion, signaled by a set curvature significantly lower than past values, highlights a potentially problematic element selection in the past. Drawing inspiration from fault-detection methodologies, we treat abrupt change in$\gamma(\mathcal{S}_{i-1}|\mathcal{A}_{i-1})$ compare to past expansion curvature as 'anomalies'. This framework allows us to view the sequential greedy process as a series of decisions, where each element added to the solution influences the overall curvature. Consequently, the trigger law is formally defined as follows.
\end{comment}
\begin{defn}[Trigger law]
Given the expansion curvature $\gamma_e(\mathcal{S}_{i}) \!\!=\!\! \max \left\{ \gamma(\mathcal{S}_0|\mathcal{A}_0),\cdots,\gamma(\mathcal{S}_{i-1}|\mathcal{A}_{i-1}) \right\}$ at stage $i$ and the set curvature at stage $i+1$, $\gamma(\mathcal{S}_{i}|\mathcal{A}_{i})$, an anomaly is considered when
$$ \gamma(\mathcal{S}_{i}|\mathcal{A}_{i}) \leq \gamma_e(\mathcal{S}_{i}).$$
\label{defn::trigger}
\boxend
\end{defn}

The trigger law checks if the current set curvature $\gamma(\mathcal{S}_{i}|\mathcal{A}_{i})$ will dominate the expansion curvature $\gamma_e(\mathcal{S}_{i})$. We expect that $\gamma(\mathcal{S}_{i}|\mathcal{A}_{i})$ increases as we more elements and diminishing return is kicking in gradually. If this condition does not hold, it suggests that an element added at iteration $\ell\leq i$ has significantly reduced the permissible expansion, indicating a potential anomaly. This trigger rule is computationally efficient, as it relies solely on the set curvatures calculated during the sequential greedy algorithm, without requiring additional queries. To illustrate, consider an example where, using sequential greedy we have reached to stage $4$ and chosen four elements. In this example suppose the expansion curvature is $\gamma_e(\mathcal{S}_3) = \max\{0,0.2, 0.5\}$, where the elements belong from left to right, respectively to $\gamma(\mathcal{S}_{0}|\mathcal{A}_{0})$, $\gamma(\mathcal{S}_{1}|\mathcal{A}_{1})$ and $\gamma(\mathcal{S}_{2}|\mathcal{A}_{2})$.  Then at $i=3$, the set curvature is $\gamma(\mathcal{S}_3|\mathcal{A}_3)=0.4$. %Here, $\gamma_e(\mathcal{S}_4)$ is dominated by the maximum set curvature up to iteration $3$ which is $0.5$. The set curvature at iteration $4$ is $0.4$.
This means that the increased path curvature of $\gamma_p(\mathcal{S}_4)\leq \gamma_{\text{e}}(\mathcal{S}_4)=\max\{0,0.2,0.5,0.4\}=0.5$ is likely related to the poor choice at stage $2$. Since $\gamma(\mathcal{S}_3|\mathcal{A}_3)=0.4 < \gamma_e(\mathcal{S}_4)=0.5$, the trigger law is activated. 

\subsection{Step-back Policy}

Once the trigger law is activated in stage $i+1$, the rewiring mechanism steps back by eliminating an element $\bar{s}$ from $\bar{\mathcal{S}}_{i+1}$ as described in line 6 of Algorithm~\ref{alg:resque_high}. The following result shows the rewiring policy is a crucial decision to ensure \textsf{ResQue Greedy} algorithm outperforms the conventional sequential greedy algorithm. 
%F bounds with respect to sequential greedy algorithm, and reducing the computational complexity.

\begin{thm}
Let $f:2^\mathcal{P}\to\mathbb{R}^{+}$ be a monotone increasing and normal submodular function over the ground set $\mathcal{P}$. Consider the sequences $\mathcal{S}_0,\cdots,\mathcal{S}_{\kappa}$
  generated by the \textsf{ResQue Greedy} algorithm (Algorithm~\ref{alg:resque_high}) and the standard sequential greedy algorithm (Algorithm~\ref{alg:sequential}). We distinguish these sequences using superscripts RSG and SG, respectively.
    The solution generated by \textsf{ResQue Greedy} always satisfies
    \begin{align*}
        f(\mathcal{S}^\text{RSG}_m) \geq  f(\mathcal{S}^\text{SG}_m)
    \end{align*}
    if and only if
    \begin{equation*}
        \gamma_p(\mathcal{S}^\text{SG}_{m-1}|s^\text{SG}_{m}) + \frac{f(\mathcal{S}^\text{RSG}_{m-1})-f(\mathcal{S}^\text{SG}_{m-1})}{f(s^\text{SG}_{m})} \geq \gamma(\mathcal{S}^\text{RSG}_{m-1}|\mathcal{A}^\text{SG}_{m-1}).
    \end{equation*}
   \label{prop:resque_greedy_vs_greedy}
\end{thm}

\begin{proof}
    Let us consider independently the sequential greedy construction as $\mathcal{S}^\text{SG}_{i+1}$ and the \textsf{ResQue Greedy} construction as $\mathcal{S}^\text{RSG}_{i+1}$. Clearly, before any rewiring is applied, $f(\mathcal{S}^\text{RSG}_{i+1}) =  f(\mathcal{S}^\text{SG}_{i+1})$. After the first rewiring operation, by definition, it is also clear that $f(\mathcal{S}^\text{RSG}_{i+1}) \geq  f(\mathcal{S}^\text{SG}_{i+1})$. Then, let us proceed to show this statement will be preserved in future iterations under certain conditions.
    
    Specifically, for any iteration $\kappa\geq m\geq i+1$ being iteration $i+1$ the stage where a rewiring was produced, by using marginal decomposition, we can establish that $f(\mathcal{S}^\text{SG}_{m}) = f(\mathcal{S}^\text{SG}_{m-1})+(1-\gamma(\mathcal{S}^\text{SG}_{m-1}|s^\text{SG}_{m})) f(s^\text{SG}_{m})$ and $f(\mathcal{S}^\text{RSG}_{m}) = f(\mathcal{S}^\text{RSG}_{m-1})+(1-\gamma(\mathcal{S}^\text{RSG}_{m-1}|s^\text{RSG}_{m})) f(s^\text{RSG}_{m})$ from where
    \begin{align*}
        f(\mathcal{S}^\text{RSG}_{m}) - f(\mathcal{S}^\text{SG}_{m}) &= f(\mathcal{S}^\text{RSG}_{m-1}) - f(\mathcal{S}^\text{SG}_{m-1}) +(1-\gamma(\mathcal{S}^\text{RSG}_{m-1}|s^\text{RSG}_{m})) f(s^\text{RSG}_{m}) -(1-\gamma(\mathcal{S}^\text{SG}_{m-1}|s^\text{SG}_{m})) f(s^\text{SG}_{m}).
    \end{align*}
    By greedy decision condition, 
    $$(1-\gamma(\mathcal{S}^\text{RSG}_{m-1}|s^\text{RSG}_{m})) f(s^\text{RSG}_{m})\geq (1-\gamma(\mathcal{S}^\text{RSG}_{m-1}|s^\text{SG}_{m})) f(s^\text{SG}_{m}).$$
    Finally, considering the definition of \emph{path curvature}, it is straightforward to see that $f(\mathcal{S}^\text{RSG}_m) \geq  f(\mathcal{S}^\text{SG}_m)$ if 
    \begin{equation*}
        \gamma_p(\mathcal{S}^\text{SG}_{m-1}|s^\text{SG}_{m}) - \gamma(\mathcal{S}^\text{RSG}_{m-1}|\mathcal{A}^\text{SG}_{m-1}) \geq -\frac{f(\mathcal{S}^\text{RSG}_{m-1})-f(\mathcal{S}^\text{SG}_{m-1})}{f(s^\text{SG}_{m})}
    \end{equation*}
    concluding the proof.
\end{proof}

Theorem~\ref{prop:resque_greedy_vs_greedy} shows that the path curvature of the truncated sequential greedy should be greater than a threshold based on the set curvature at each iteration of the sequential greedy algorithm after the rewiring in order to achieve a better performance with the proposed \textsf{ResQue Greedy}. This observation directly implies the urge of reducing the \emph{set curvature} at each stage as an anomaly is detected. 

Therefore, when an anomaly detected, the rewiring policy seeks to remove the element $s_{\ell}$ added in the iteration $\ell\leq i$ being $\ell$ the stage that dominates among the rest of computed set curvatures on the expansion curvature. Considering the previous example, let us suppose $\gamma_e(\mathcal{S}_4)=\max\{0,0.2,0.5,0.4\}$, then the expansion curvature is $0.5$, governed by $i=3$ which represents the diminished caused by the addition of $s_2$, being $\bar{s}=s_2$ the candidate to be removed. 

\begin{defn}[Step-back Policy]
    Given the expansion curvature $\gamma_e(\mathcal{S}_{i+1})=\left\{ \gamma(\mathcal{S}_{\ell-1}|\mathcal{A}_{\ell-1}) \right\}_{\ell=1,\cdots,i+1}$ at stage $i+1$ when the trigger law is activated, the discarded element is $s_{j-1}\in\bar{\mathcal{S}}_{i+1}$ where $s_{j-1}$ accomplishes that $$j=\text{argmax}_{\ell\in\{1,\cdots,i+1\}}\left\{\gamma(\mathcal{S}_{\ell-1}|\mathcal{A}_{\ell-1})\right\}_{\ell=1,\cdots,i+1}.$$ \boxend
    \label{defn::policy}
\end{defn}
\smallskip

After the step-back is produced, the set curvature at each stage will be reduced accordingly to the removed element, directly implying a reduction on the expansion curvature. In order to avoid recomputing the set curvature for each stage of the new solution $\mathcal{S}^\prime_i$, we will recall on the fact that, before the step-back, only sets $\mathcal{S}_\ell$ where $\ell\geq j$ contained $\bar{s}=s_{j-1}$. Therefore, only $\gamma(\mathcal{S}_\ell|\mathcal{A}_\ell)$ for  $\ell\geq j$ must be recomputed. Additionally, by Lemma~\ref{lem::properties}, it is known that after the step-back and removing stage $j$, $\gamma(\mathcal{S}^\prime_\ell|\mathcal{A}^\prime_\ell)\in[\gamma(\mathcal{S}_{\ell+1}|\mathcal{A}_{\ell+1}),\gamma(\mathcal{S}_{\ell-1}|\mathcal{A}_{\ell-1})]$. Consequently, we propose an heuristic update by considering the mean value on the expected interval such that
\begin{equation}
    \label{eqn::update_rule}\gamma(\mathcal{S}^\prime_\ell|\mathcal{A}^\prime_\ell)\! = \! \gamma(\mathcal{S}_{\ell\!-\!1}|\mathcal{A}_{\ell\!-\!1}) + \frac{\gamma(\mathcal{S}_{\ell\!+\!1}|\mathcal{A}_{\ell\!+\!1})\!-\!\gamma(\mathcal{S}_{\ell\!-\!1}|\mathcal{A}_{\ell\!-\!1})}{2}.
\end{equation}

To exemplify this update rule, let us continue with the previous example where $\gamma_e(\mathcal{S}_4)=\max\{0,0.2,0.5,0.4\}$ and $\bar{s}=s_2$. Then, $\gamma_e(\mathcal{S}^\prime_3)=\max\{0,0.2,0.3\}$ where, $0$ and $0.2$ are not effected by the removing of $s_2$, $0.5$ is removed with $s_2$ and $0.4$ turns to be $\gamma(\mathcal{S}^\prime_2|\mathcal{A}^\prime_2) = 0.2 + \frac{1}{2}(0.4-0.2) = 0.3$.

%%%%%%%%%%%%%%%%%%%%%%%%%%%%%%%%%%%%%%%%%%%%%%%%%%%%%%%%%%%%%%%%%%%%%%%%%%%%%%%%%%%%%%%%%%%%%%%%

\section{Numerical Example}
\label{sec:numerical}
We demonstrate our results via two coverage problems, a multi-sensor deployment and multi-agent monitoring for space exploration. In these problems, a set of $m$ information points, denoted by $\mathcal{V}$, are scattered in an environment. The objective is to cover as many points from $\mathcal{V}$ as possible by deploying $\kappa$ assets with circular coverage footprint in a set of pre-specified deployment points. The objective is accomplished by solving a maximization problem of the form~\eqref{eq::mainProblem} with $\kappa$ and the utility function given by $ f(\mathcal{S}) = \sum\nolimits_{p \in \mathcal{V}} g(p)$, where $g(p)=1$ if there exists at least one element $\ell \in \mathcal{S}$ such that $\|c - p\| \leq \rho_{\ell}$; otherwise $g(p)=0$. Here, $\rho_{\ell}$ is the coverage radius of sensor $\ell$. This utility function is known to be submodular and monotone increasing~\cite{MS-AK-XS-DS-TC:18}.

\begin{figure}[t]
   \centering
        \includegraphics[width=0.45\textwidth]{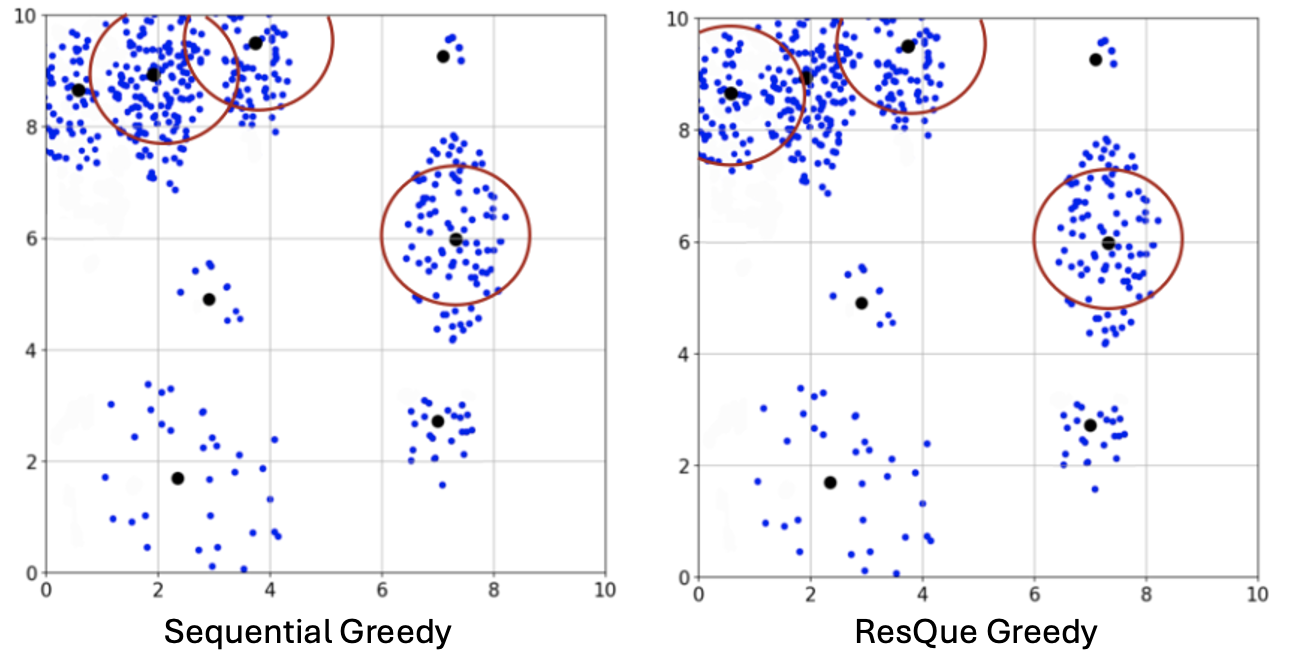}
    \caption{{\small A multi-sensor deployment problem: the possible allocation points are shown by bigger black dots, while the information points are shown by blue dots. In this simple case, we can see that \textsf{ResQue Greedy}'s rewiring at the last step by dropping the second deployment location chosen by the sequential greedy algorithm and retrying to choose another element leads to an improvement: the sequential greedy solution obtains a coverage of $520$ features collected, whereas \textsf{ResQue Greedy} yields a greater coverage of $657$ features.}}
    \label{fig::environment}
\end{figure}
\begin{comment}
In {\begin{tikzpicture}[
            scale=0.6,
            every node/.style={draw, circle, inner sep=1pt, font=\footnotesize},
            level 1/.style={sibling distance=4cm},
            level 2/.style={sibling distance=1.5cm},
            level 3/.style={sibling distance=0.8cm}
        ] \draw[fill=blue](0,0) circle (2pt); \end{tikzpicture}} the points of interest $\mathcal{B}$ and in {\begin{tikzpicture}[
            scale=0.6,
            every node/.style={draw, circle, inner sep=1pt, font=\footnotesize},
            level 1/.style={sibling distance=4cm},
            level 2/.style={sibling distance=1.5cm},
            level 3/.style={sibling distance=0.8cm}
        ] \draw[fill=black](0,0) circle (3pt); \end{tikzpicture}} the possible allocation points $\mathcal{P}$. 
\end{comment}

\begin{figure}
    \centering
    \includegraphics[width=0.35\textwidth]{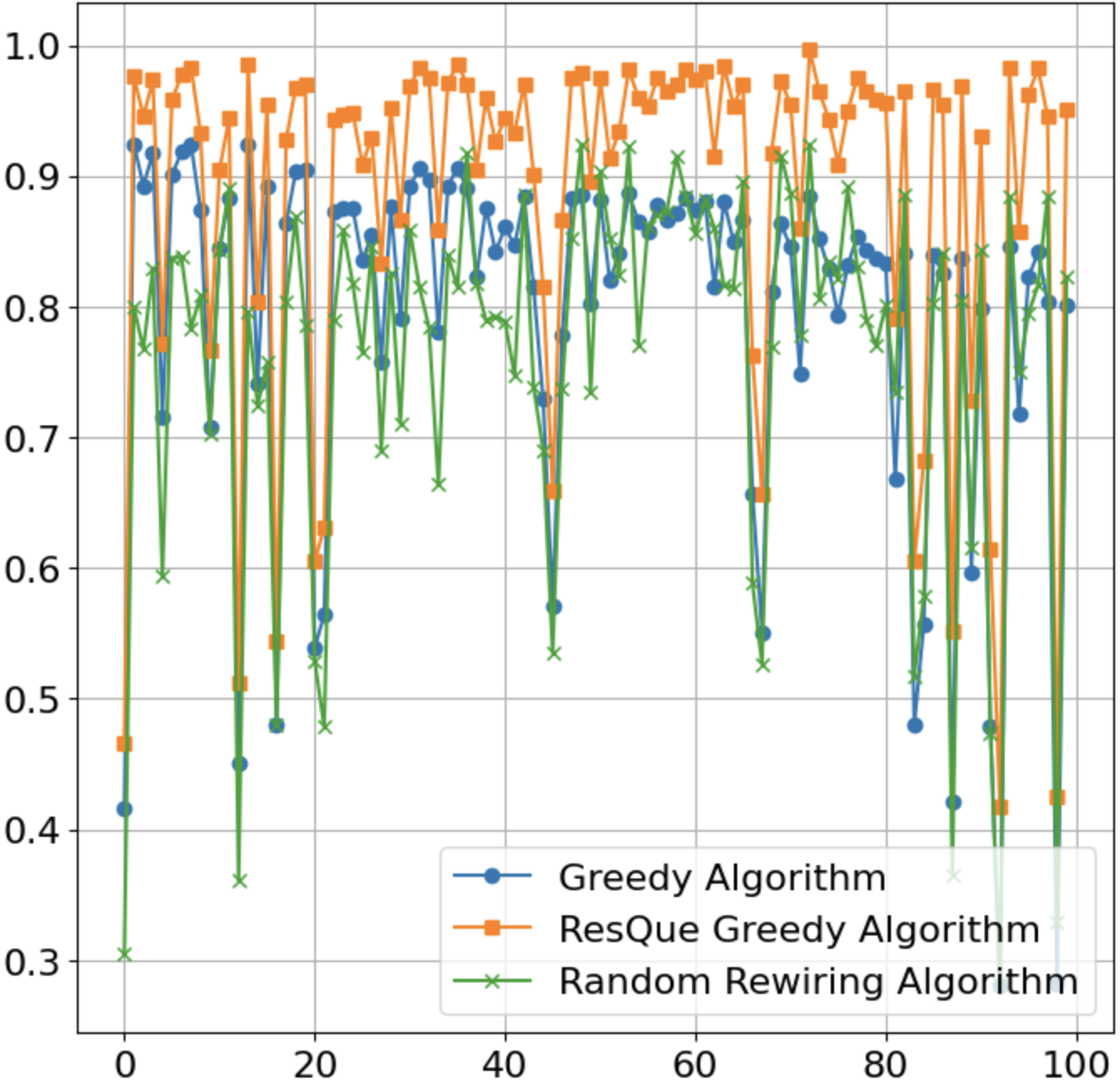}
    \caption{Multi-sensor deployment: $100$ Monte-Carlo simulation.}
    \label{fig:monte_carlo}
\end{figure}

\emph{Multi-sensor deployment problem}: A set of $m$ information source points are randomly generated by sampling from a Gaussian Mixture Model. The set of these points is denoted by $\mathcal{V}$.  A simple small scale version of the problem is illustrated in Fig.~\ref{fig::environment}. We implement a Monte-Carlo simulation to generate $100$ different random environments. The problem dimensions have been randomly chosen in order to consider different magnitudes of the space. The order of the random variables are as follows: $10 - 50$ prespecified allowable allocation points (i.e., $|\mathcal{P}|$ is $5-50$), and  $5,000 - 25,000$ information points distributed considering a random curvature ratio between $0 - 1$. The number of deployed sensors in each simulation is between $5$ to $0.8\times |\mathcal{P}|$. The sensors allocation points $\mathcal{P}$ have been generated by considering a Determinantal Point Process~\cite{RI-JB:15} in order to create different orders of submodularity on the problem (different values of curvature $c$). The results shown in Table~\ref{tab:results} and Fig.~\ref{fig:monte_carlo} shed light in the performance improvement of the proposed strategy. \textsf{ResQue Greedy} Algorithm out-performs in average the Sequential Greedy approach and, besides, as stated previously, it never under-performs the classical bounds under any situation, which clearly demonstrates its practical usage as an opportunistic improvement. It is noted that, even the increase in the computational time, the proposed method is still polynomial-time with also an small increase in the query time. Both algorithms have also been contrasted with a random rewiring strategy without a trigger mechanism. This comparison evidences the fact that there must exist certain strategy in the rewiring process in order to take advantage of it. % and not guiding the solution to even worser paths or exponentially increase the computational time. 
Finally, it must be observed that the \textsf{ResQue Greedy} improvement is not substantial in some situations. This could be attributed to problems where the function has already a low total curvature and the function already behaves well using the sequential greedy algorithm and the trigger law is activated only few times.
%during the process. This exposes the importance on the trigger law and the fact that an aggressive rule can potentiate the benefits of the rewiring methods. 

\begin{table}[t]
    \caption{{\small Average normalized (with respect to total number of information points in each simulation) coverage results of Monte-Carlo simulations for the first example.}}
      \centering
    \begin{tabular}{|c|c|c|c|}
        \hline
          &  \textbf{Coverage} & \textbf{Time}~[s] & \textbf{Queries}\\
        \hline
        \textsf{Greedy} & 0.7823 & 0.7749 & 158.80\\
        \hline
        \textsf{ResQue Greedy} & 0.8874 & 0.8507 & 187.58\\
        \hline
        \textsf{Random Rewiring Greedy} & 0.7439 & 2.3034 & 257.97\\
        \hline
    \end{tabular}
    \label{tab:results}
\end{table}

\begin{figure}
    \centering
    \includegraphics[width=0.9\linewidth]{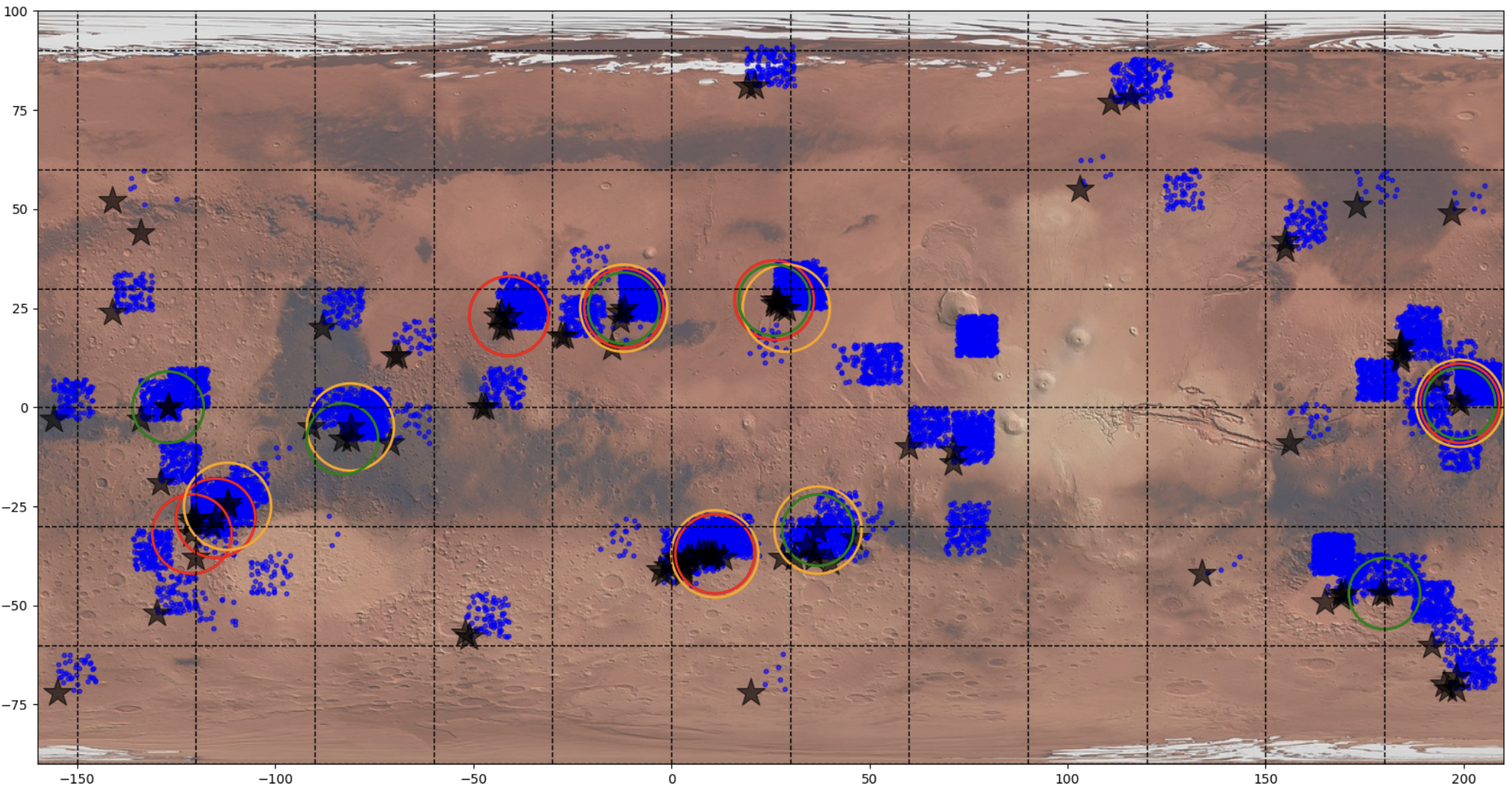}
    \caption{{\small Feature points (blue dots) and potential deployment points (black $\star$). Agent deployments: standard sequential greedy (red), $\textsf{ResQue Greedy}$ allocations (orange), and random rewiring greedy allocations (green). All sensors are homogenous with the same circular range. To visualize overlapping deployments using our three different algorithms, sensor ranges are depicted with slightly varying radii for each algorithm. Note the $\textsf{ResQue Greedy}$ resolves some of the overlapped deployments, such as those two in the lower left corner. %reallocation of the top-left agent, originally selected by the sequential greedy algorithm, to a feature-richer area.
    }}
    \label{fig:mars}
\end{figure}

%%%%%%%%%%%%%%%%%%%%%%%%%%%%%%%%%%%%%%%%%%%%%%%%%%%%%%%%%%%%%%%%%%%%%%%%%%%%%%%%%%%%%%%%%%%%%%%%%%%%%%%%%%%%%%%%%%%%%%%%%%%%%%%%%%%%%%%%%%%%%%%%%%%%%%%%%%%%%%%%%%%%%%%%%%%%%%%%%%%%%%%%%%
\emph{Multi-agent monitoring for space exploration}: As a second example, we consider a real-world problem in an space exploration setting. %Specifically, we consider the challenge of optimizing exploration routes for spatial missions, particularly in the context of exploring unknown planets. 
In such missions, it is crucial to use resources efficiently to maximize the discovery of valuable features. Since spatial robots, such as rovers, require careful management of limited energy and are costly to deploy, optimizing their exploration routes is essential. This problem is fundamentally a coverage problem and can be formulated as Problem~\eqref{eq::mainProblem}. Given a set of potential landing points $\mathcal{P}$, the goal is to direct missions to those locations that allow for the coverage of the most features $\mathcal{V}$, thereby minimizing energy consumption and reducing overall mission costs. In this work, we utilize the DoMars16k dataset~\cite{TW-MG-JP-TS-TW-KW-CW:20}, which includes $132$ potential allocation points for gathering up to $16,150$ interesting surface features, essential for planning efficient land rover trajectories, see Fig.~\ref{fig:mars}. By allowing for the deployment of up to $\kappa = 7$ rovers, the optimal landing sites comparing the \textsf{ResQue Greedy}, the Sequential Greedy and the random rewiring algorithm are compared in Table~\ref{tab:results2}. As we can see, with a slight increase in computation, we can lead to a tighter sub-optimal solution which can increase the effectiveness of the deployment. % with, as mentioned, a reduction on the overall costs of the missions. 

\begin{comment}
    Planar global mosaic of the Martian surface ($925 m/px$, Robinson projection, planetocentric adapted from~\cite{MarsVikingMosaic}). In {\begin{tikzpicture}[
            scale=0.6,
            every node/.style={draw, circle, inner sep=1pt, font=\footnotesize},
            level 1/.style={sibling distance=4cm},
            level 2/.style={sibling distance=1.5cm},
            level 3/.style={sibling distance=0.8cm}
        ] \draw[fill=blue](0,0) circle (2pt); \end{tikzpicture}} the interesting features of the surface and the {\begin{tikzpicture}[
        scale=0.6,
        every node/.style={draw, star, star points=5, inner sep=1pt, font=\footnotesize},
        level 1/.style={sibling distance=4cm},
        level 2/.style={sibling distance=1.5cm},
        level 3/.style={sibling distance=0.8cm}]
        \draw[fill=black](0,0) node[star, star points=5, fill=black, minimum size=5pt] {};
      \end{tikzpicture}} stand for the prospective landing sites for different missions such as ExoMars or Mars2020. In blue, the greedy selections, in orange the $\textsf{ResQue Greedy}$ allocations and in green, the random rewiring greedy implementation.
\end{comment}

\begin{table}[t]
    \centering
    \caption{{\small Results  for the second problem.} }
    \begin{tabular}{|c|c|c|c|}
        \hline
          &  \textbf{Coverage} & \textbf{Time} [s] & \textbf{Queries}\\
        \hline
        \textsf{Greedy} & $6,658$ & $0.0974$ & $721$\\
        \hline
        \textsf{ResQue Greedy} & $7,714$ & $0.1082$ & $825$\\
        \hline
        \textsf{Random Rewiring Greedy} & $6,853$ & $0.1002$ & $762$\\
        \hline
    \end{tabular}
    \label{tab:results2}
\end{table}

\section{Conclusion and Future Directions}
\label{sec:conclusions}
This paper introduced \textsf{ResQue Greedy}, a novel framework that enhances the sequential greedy algorithm for submodular maximization under cardinality constraints. By employing a curvature-aware rewiring strategy, \textsf{ResQue Greedy} dynamically adjusts the solution path, improving approximation performance without significant computational overhead. Numerical experiments confirmed its effectiveness in achieving tighter near-optimality bounds. The core contribution is the trigger law, which utilizes set and expansion curvatures to efficiently determine when rewiring is beneficial. This approach offers a practical method for improving solution quality in resource allocation problems. Future research directions include extending \textsf{ResQue Greedy} to handle more complex constraints like matroid or knapsack constraints. Investigating adaptive strategies within the trigger law. Exploring tighter theoretical bounds and applying \textsf{ResQue Greedy} to real-world large-scale problems also warrant further investigation.

\bibliographystyle{ieeetr}
\bibliography{main}

\end{document}